\documentclass[preprint,showpacs,showkeys,preprintnumbers,amsmath,amssymb,aps]{revtex4-1}

\usepackage{bm,amssymb,amsmath,mathrsfs}
\usepackage[usenames]{color}

\usepackage{dcolumn}
\usepackage[pdftex]{graphicx}

\newcommand{\bc}{\begin{center}}
\newcommand{\ec}{\end{center}}
\newcommand{\bit}{\begin{itemize}}
\newcommand{\eit}{\end{itemize}}
\newcommand{\bq}{\begin{equation}}
\newcommand{\eq}{\end{equation}}

\begin{document}

\title{Equation for spin decoherence rate in an all-electric ring}

\author{S.~R.~Mane}
\email{srmane001@gmail.com}

\affiliation{Convergent Computing Inc., P.~O.~Box 561, Shoreham, NY 11786, USA}

\begin{abstract}
There is a quantitative error in the derivation for the spin decoherence rate by Talman in IPAC2012.
The crucial point is a subtle confusion between the concept of `longitudinal' as in `along the reference orbit' and `parallel to the particle velocity.'
They are {\em not} the same direction, and the distinction is significant
for high-precision experiments to search for a possible nonzero
electric dipole moment (EDM) of a charged particle in a storage ring.
\end{abstract}



\maketitle

\section{Introduction}
Richard and John Talman published a paper in IPAC2012 \cite{TalmanIPAC2012}
giving some details of the orbital and spin motion in an all-electric storage ring.
There is a quantitative error in Talman's derivation.
The core of the matter is a subtle but significant detail: the concept of `longitudinal.'
There is a confusion between `longitudinal' as in `along the reference orbit' and `longitudinal' as in `parallel to the particle velocity.'
They are {\em not} the same direction, and the distinction is significant
for high-precision experiments to search for a possible nonzero
electric dipole moment (EDM) of a charged particle in a storage ring.

\section{Basic notation}
I shall treat a particle of mass $m$, charge $e$,
with velocity $\bm{v}=\bm{\beta}c$ and Lorentz factor $\gamma=1/\sqrt{1-\beta^2}$.
(Talman \cite{TalmanIPAC2012} denotes the particle mass by $m_p$,
and I shall also use this notation below.)
The particle spin $s$ is treated as a unit vector and $g$ denotes the particle $g$-factor.
There is no magnetic field in the model, the ring is all-electric.
For simplicity of the exposition, I shall treat a smooth focusing model below.
I employ cylindrical polar coordinates $(r,\theta,z)$.
Following Talman \cite{TalmanIPAC2012},
I shall treat orbital and spin motion in the horizontal plane only.
In the horizontal plane, the electric field points radially.

\section{Helicity: angle $\alpha$}
First, I need to define the important angle $\alpha$,
which is the angle between the spin unit vector $\bm{s}$ and
the unit vector in the direction of the velocity be $\hat{\bm{\beta}}$.
(Both vectors are assumed to lie in the horizontal plane, as stated above.)
Following Talman \cite{TalmanIPAC2012},
we go counterclocksise from $\hat{\bm{\beta}}$ to $\bm{s}$.
Then
\bq
\bm{s}\cdot\hat{\bm{\beta}} = \cos\alpha \,,\qquad
\bm{s}\times\hat{\bm{\beta}} = -\sin\alpha\,\hat{\bm{z}} \,.
\eq
Also define $x$ via $r = r_0+x$, where $r_0$ is a reference radius.
Talman defines the electric field to point radially {\em inward},
so $\bm{E} = -E(x)\,\hat{\bm{r}}$.
Then Talman writes (eq.~(10) in \cite{TalmanIPAC2012})
\bq
\label{eq:talmaneq10}
\frac{d\alpha}{dt} = \frac{eE(x)}{m_pc}\,\biggl(\frac{g\beta(x)}{2}-\frac{1}{\beta(x)}\biggr) \,.
\eq
Talman \cite{TalmanIPAC2012} cites
Jackson \cite{JacksonCE3} for the above equation.
(Note that all references to Jackson's textbook in this note are to the second edition.) 
However Jackson's textbook does {\em not} contain the above equation.
It has been {\em derived} from an equation in Jackson's text.
We must therefore begin our analysis from the actual equation written by Jackson.

\section{Equation from Jackson}
Jackson \cite{JacksonCE3} 
writes the following equation for the evolution of the longitudinal spin component
\bq
\label{eq:jdjeq11.171}
\frac{d\ }{dt}(\hat{\bm{\beta}}\cdot\bm{s}) = -\frac{e}{mc}\,\bm{s}_\perp\cdot\biggl[
\biggl(\frac{g}{2}-1\biggr)\,\hat{\bm{\beta}}\times\bm{B} + \biggl(\frac{g\beta}{2}-\frac{1}{\beta}\biggr)\,\bm{E}\,\biggr]\,.
\eq
Note that this equation is {\em coordinate-free}, and is {\em completely general},
and is {\em not} restricted to motion in a plane.
Here $\bm{E}$ and $\bm{B}$ are the external electric and magnetic fields, respectively, 
and $\bm{s}_\perp$ is the spin component orthogonal to the direction of the velocity $\hat{\bm{\beta}}$, viz.
\bq
\bm{s}_\parallel = \bm{s}\cdot\hat{\bm{\beta}}\,\hat{\bm{\beta}} \,,
\qquad
\bm{s}_\perp = \bm{s} - \bm{s}_\parallel
= \bm{s} - \bm{s}\cdot\hat{\bm{\beta}}\,\hat{\bm{\beta}} 
= \hat{\bm{\beta}} \times (\bm{s} \times \hat{\bm{\beta}}) \,.
\eq
To make contact with eq.~\eqref{eq:talmaneq10}, 
we restrict the motion to the horizontal plane.
As stated previously, there is no magnetic field in our model
and the electric field in the horizontal plane is radial $\bm{E}=-E(x)\hat{\bm{r}}$.
Then
\begin{align}
\frac{d\ }{dt}(\hat{\bm{\beta}}\cdot\bm{s}) &= \frac{eE(x)}{m_pc}\,
\biggl(\frac{g\beta}{2}-\frac{1}{\beta}\biggr)
(\hat{\bm{\beta}} \times (\bm{s} \times \hat{\bm{\beta}})) \cdot \hat{\bm{r}} 
\\
\sin\alpha\,\frac{d\alpha}{dt} &= \frac{eE(x)}{m_pc}\,
\biggl(\frac{g\beta}{2}-\frac{1}{\beta}\biggr)
(\hat{\bm{\beta}} \times \hat{\bm{z}}\, \sin\alpha) \cdot \hat{\bm{r}} 
\\
\label{eq:dadtfromjdj}
\frac{d\alpha}{dt} &= 
\frac{eE(x)}{m_pc}\, \biggl(\frac{g\beta}{2}-\frac{1}{\beta}\biggr)
{\color{red} \hat{\bm{\beta}} \cdot \hat{\bm{\theta}}} \,.
\end{align}
The final factor of $\hat{\bm{\beta}} \cdot \hat{\bm{\theta}}$
is absent from Talman's IPAC2012 paper \cite{TalmanIPAC2012} 
(see eq.~\eqref{eq:talmaneq10} above).
This is the subtlety of the concept of `longitudinal,'
viz.~the direction along the reference orbit is {\em not}
the same as the direction of the particle velocity.
This has consequences which I shall spell out below.

\section{Equation for $d\alpha/d\theta$}
Starting from the equation for $d\alpha/dt$,
Talman \cite{TalmanIPAC2012} derives an equation for $d\alpha/d\theta$
(eq.~(12) in \cite{TalmanIPAC2012}).
An equation for $d\alpha/d\theta$, 
essentially the rate of change of $\alpha$ per turn around the ring,
is more useful for accelerator physics work.
I derive the equation for $d\alpha/d\theta$ as follows.
Note that
\bq
\hat{\bm{\beta}} \cdot \hat{\bm{\theta}}
= \frac{v_\theta}{\beta c} = \frac{r}{\beta c}\,\frac{d\theta}{dt} \,.
\eq
Hence from eq.~\eqref{eq:dadtfromjdj},
\begin{align}
\frac{d\alpha}{dt} &= 
\frac{eE(x)}{m_pc}\, \biggl(\frac{g\beta}{2}-\frac{1}{\beta}\biggr)\,\frac{r}{\beta c}\,\frac{d\theta}{dt}
\\
\label{eq:dadtheta_me}
\frac{d\alpha}{d\theta} &= 
\frac{eE(x)r}{m_pc^2\beta}\, \biggl(\frac{g\beta}{2}-\frac{1}{\beta}\biggr) \,.
\end{align}
Talman's derivation to obtain an equation for $d\alpha/d\theta$
proceeds as follows.
Talman notes that the (vertical) angular momentum $L$ is conserved, 
since the force (electric field) is radial, 
hence (eq.~(11) in \cite{TalmanIPAC2012})
\bq
\frac{d\theta}{dt} = \frac{L}{\gamma m_pr^2} \,.
\eq
Hence from eq.~\eqref{eq:talmaneq10}
\bq
\label{eq:dadtheta_talman}
\begin{split}
\frac{d\alpha}{d\theta} &= \frac{eE(x){\color{red} \bm{r^2\gamma}}}{Lc}
\,\biggl(\frac{g\beta}{2}-\frac{1}{\beta}\biggr) 
\\
&= \frac{eE(x)(r_0+x)^2}{Lc\beta}\,\biggl(\frac{g}{2}\,\beta^2\gamma - \gamma\biggr) 
\\
&= \frac{eE(x)(r_0+x)^2}{Lc\beta}\,\biggl(\Bigl(\frac{g}{2}-1\Bigr)\gamma-\frac{g/2}{\gamma}\biggr) \,.
\end{split}
\eq
This is eq.~(12) in \cite{TalmanIPAC2012}.

\section{Differences/consequences}
Comparing the two expressions,
I have $E(x){\color{red} \bm{r/\beta}}$ (eq.~\eqref{eq:dadtheta_me})
whereas Talman has $E(x){\color{red} \bm{r^2\gamma}}$ (eq.~\eqref{eq:dadtheta_talman})
There are of course also factors of $L$, $m_p$ and $c$ to balance the dimensions, but they are constants.
The different dependence on the orbit is sufficient to yield noticeable quantitative differences in a high-precision analysis.
The distinction between the direction along the reference orbit 
and the direction of the particle velocity is subtle but significant.


\vfill\pagebreak
\appendix
\section{Averages over the orbit}
Talman \cite{TalmanIPAC2012}
published the following formula for the spin decoherence rate 
for orbital and spin motion in the horizontal plane in an all-electric ring
(eq.~(17) in \cite{TalmanIPAC2012})
\bq
\label{eq:talmanipac2012_eq17}
-\biggl\langle \frac{d\alpha}{d\theta} \biggr\rangle \approx \frac{E_0r_0\gamma_0}{(p_0c/e)\beta_0}\,
\biggl(\biggl\langle \frac{\gamma}{\gamma_0} -1 \biggr\rangle 
+m\,\biggl\langle \frac{x}{r_0} \biggr\rangle 
-\frac{m^2-m}{2}\,\biggl\langle \frac{x^2}{r_0^2} \biggr\rangle \biggr)\,.
\eq
Talman employs the notation $m$ for the field index (see below) and uses $m_p$ for the particle mass.
There is an error of algebra in the above formula, in addition to other errors  I have pointed out above.
Talman eq.~(13) states
\bq
\label{eq:talmanipac2012_eq13}
\biggl\langle \frac{d\alpha}{d\theta} \biggr\rangle \approx \biggl\langle\frac{eE_0(r_0+x)^2}{(Lc\beta(x)}\biggr\rangle\,
\biggl(\Bigl(\frac{g}{2}-1\Bigr)\langle \gamma\rangle -\frac{g}{2}\Bigl\langle\frac{1}{\gamma}\Bigr\rangle \biggr) \,.
\eq
Next Talman employs the relativistic virial theorem to deduce (Talman eq.~(16))
\bq
\biggl\langle\frac{1}{\gamma}\biggr\rangle = \langle\gamma\rangle - \frac{E_0r_0}{m_pc^2/e}\,\biggl\langle\frac{r_0^m}{r^m}\biggr\rangle \,.
\eq
Use this in eq.~\eqref{eq:talmanipac2012_eq13}.
Also we operate at the magic gamma, so $a=1/(\beta_0^2\gamma_0^2)$ and $g/2 = 1+a = 1/\beta_0^2$.
Then
\bq
\label{eq:talmanipac2012_usevirial}
\begin{split}
-\biggl\langle \frac{d\alpha}{d\theta} \biggr\rangle &\approx \biggl\langle\frac{eE_0(r_0+x)^2}{(Lc\beta(x)}\biggr\rangle\,
\biggl( -\Bigl(\frac{g}{2}-1\Bigr)\langle \gamma\rangle 
+\frac{g}{2}\,\langle\gamma\rangle 
-\frac{g}{2}\,\frac{E_0r_0}{m_pc^2/e}\,\biggl\langle\frac{r_0^m}{r^m}\biggr\rangle 
\biggr) 
\\
&\simeq \frac{eE_0r_0^2}{p_0r_0c\beta_0}\,
\biggl( \langle \gamma\rangle 
-\frac{1}{\beta_0^2}\,\frac{m_pc^2\gamma_0\beta_0^2}{m_pc^2}\,\biggl\langle\frac{r_0^m}{r^m}\biggr\rangle 
\biggr) 
\\
&\simeq \frac{E_0r_0\gamma_0}{(p_0c/e)\beta_0}\,
\biggl( \biggl\langle \frac{\gamma}{\gamma_0} \biggr\rangle 
-\biggl\langle\frac{r_0^m}{r^m}\biggr\rangle 
\biggr) 
\\
&\simeq \frac{E_0r_0\gamma_0}{(p_0c/e)\beta_0}\,
\biggl( \biggl\langle \frac{\gamma}{\gamma_0} \biggr\rangle 
-\biggl\langle 1 - m\,\frac{x}{r_0} + \frac{m(1+m)}{2}\,\frac{x^2}{r_0^2} \biggr\rangle 
\biggr) 
\\
&= \frac{E_0r_0\gamma_0}{(p_0c/e)\beta_0}\,
\biggl( \biggl\langle \frac{\gamma}{\gamma_0} - 1 \biggr\rangle 
+m\, \biggl\langle \frac{x}{r_0}  \biggr\rangle 
-\frac{\color{red}\bm{m^2+m}}{2}\, \biggl\langle \frac{x^2}{r_0^2} \biggr\rangle 
\biggr) \,.
\end{split}
\eq
Hence there is an error of algebra in the last term of Talman eq.~(17)
(see eq.~\eqref{eq:talmanipac2012_eq17});
the coefficient should be $m^2+m$ not $m^2-m$.

\end{document}